\documentclass[pdflatex,sn-mathphys-num]{sn-jnl}

\usepackage{graphicx}%
\usepackage{multirow}%
\usepackage{amsmath,amssymb,amsfonts}%
\usepackage{amsthm}%
\usepackage{mathrsfs}%
\usepackage[title]{appendix}%
\usepackage{xcolor}%
\usepackage{textcomp}%
\usepackage{manyfoot}%
\usepackage{booktabs}%
\usepackage{algorithm}%
\usepackage{algorithmicx}%
\usepackage{algpseudocode}%
\usepackage{listings}%
\usepackage{array}
\usepackage{makecell}
\usepackage{colortbl}
\usepackage{longtable}
\usepackage{adjustbox}
\usepackage{pdflscape}
\usepackage{placeins}

\theoremstyle{thmstyleone}%
\theoremstyle{thmstyletwo}%

\theoremstyle{thmstylethree}%

\begin{document}
\title[Article Title]{A 103-TOPS/mm$^2$ Integrated Photonic Computing Engine Enabling Next-Generation Reservoir Computing}


\author[1]{\fnm{Dongliang} \sur{Wang}}\email{1155167615@link.cuhk.edu.hk}

\author[1]{\fnm{Yikun} \sur{Nie}}\email{yikun.nie@link.cuhk.edu.hk}

\author[1]{\fnm{Gaolei} \sur{Hu}}\email{gaoleih@link.cuhk.edu.hk}

\author[1]{\fnm{Hon Ki} \sur{Tsang}}\email{hktsang@ee.cuhk.edu.hk}

\author*[1]{\fnm{Chaoran} \sur{Huang}}\email{crhuang@ee.cuhk.edu.hk}

\affil*[1]{\orgdiv{Department of Electronic Engineering}, \orgname{The Chinese University of Hong Kong}, \orgaddress{\street{Shatin}, \city{Hong Kong SAR}, \country{ China}}}


\abstract{Reservoir computing (RC) is a leading machine learning algorithm for information processing due to its rich expressiveness. A new RC paradigm has recently emerged, showcasing superior performance and delivering more interpretable results with shorter training data sets and training times, representing the next generation of RC computing. This work presents the first realization of a high-speed next-generation RC system on an integrated photonic chip. Our experimental results demonstrate state-of-the-art forecasting and classification performances under various machine learning tasks and achieve the fastest speeds of 60 Gbaud and a computing density of 103 tera operations/second/mm$^2$ (TOPS/mm$^2$). The passive system, composed of a simple star coupler with on-chip delay lines, offers several advantages over traditional RC systems, including no speed limitations, compact footprint, extremely high fabrication error tolerance, fewer metaparameters, and greater interpretability. This work lays the foundation for ultrafast on-chip photonic RC, representing significant progress toward developing next-generation high-speed photonic computing and signal processing. }




\maketitle

\section{Introduction}\label{sec1}

Artificial intelligence (AI) is one of the most transformative and disruptive technologies in the 21st century. The rapid growth in data volume facilitates the development of larger and more complex neural network models, such as ChatGPT, Gemini, and Sora. However, this trend also presents significant challenges in existing electronic computing hardware, particularly in terms of both computing speed and power consumption~\cite{furber2016large, liu2024sora}. These challenges have become a critical bottleneck for the advancement of AI, and drive the exploration of innovative neural network computing hardware and novel concepts. Photonics is increasingly appealing for implementing neural networks due to its outstanding potential in accelerating processing speed, reducing operation power and latency, surpassing the capabilities of traditional electronics. Many innovative optical neural network architectures, using on-chip integrated photonics, optoelectronics, diffractive optics and so on, have been demonstrated~\cite{shen2017deep,lin2018all, feldmann2019all, xu202111, shastri2021photonics, huang2021silicon, feldmann2021parallel, xu2022self, wang2023image, xu2024large, dong2023higher,ashtiani2022chip, chen2023all, liu2022programmable}.

\begin{figure}[h]
\centerline{\includegraphics[scale=0.8]{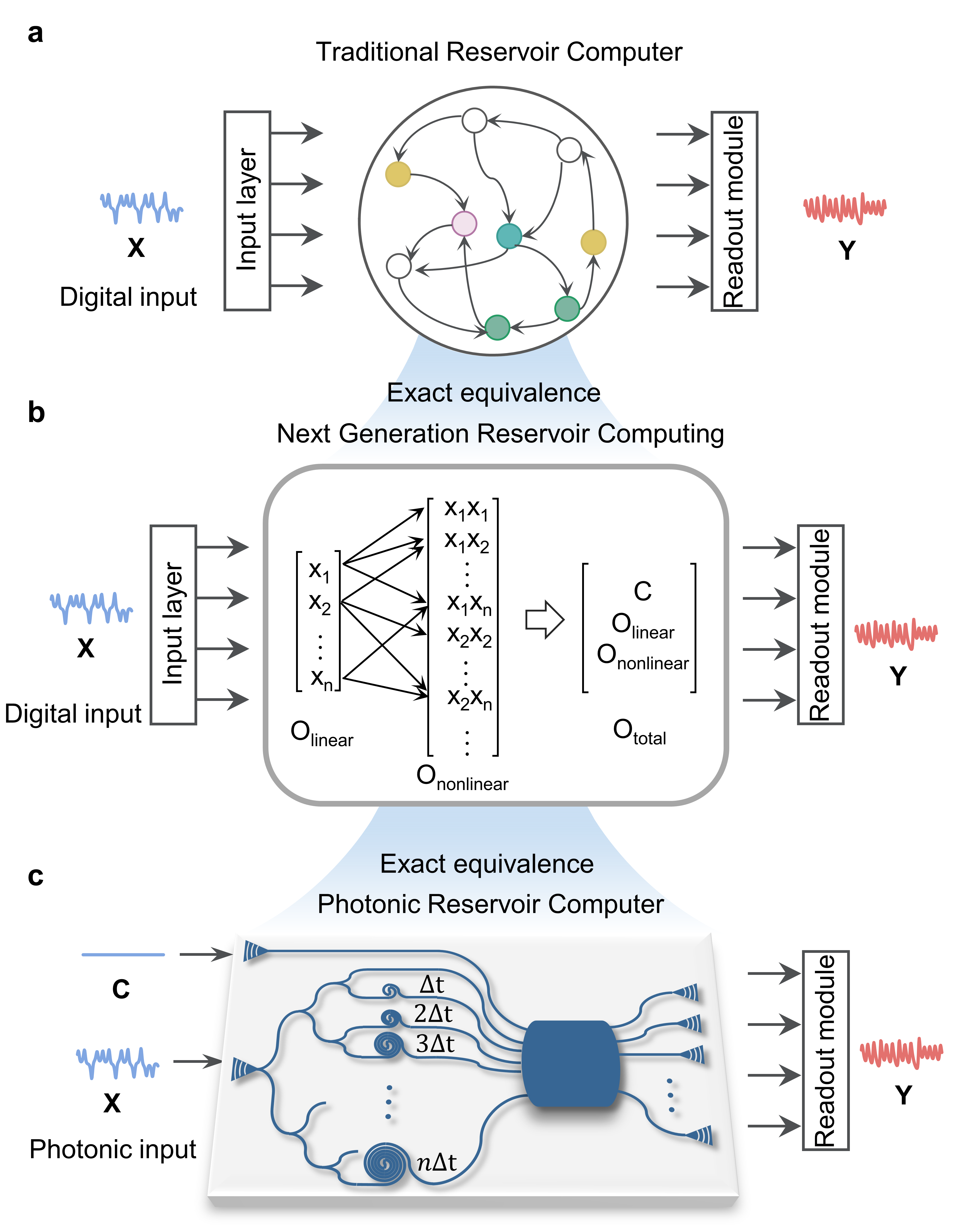}}
\caption{\textbf{Basic principle of reservoir computing. a,} The framework of traditional reservoir computing systems, including the input layer, the reservoir layer, and the readout layer. \textbf{b,} The flow of next-generation RC, which is exactly equivalent to traditional RC. NG-RC employs a constant $C$, input data $O_{linear}$, and the quadratic functionals of input data $O_{nonlinear}$ to generate its output. \textbf{c,} The schematic diagram of our photonic RC, which is perfectly equivalent to NG-RC. The proposed photonic reservoir computing is realized using on-chip delay lines and a star coupler. }
\label{fig1}
\end{figure}

Photonic reservoir computing (RC) is currently receiving considerable research interest among various photonic neural network architectures, because it avoids the burden of delicate training and configuring every optical device to realize the optimal weights~\cite{paquot2012optoelectronic, larger2012photonic, brunner2013parallel, vandoorne2014experimental, vinckier2015high,larger2017high, antonik2019human, rafayelyan2020large, sunada2021photonic, nakajima2021scalable, lupo2023deep, shen2023deep}. RC is derived from a recurrent artificial neural network, with a three-layer architecture: an input layer, a reservoir layer, and a readout layer~\cite{yan2024emerging}. The input layer receives information and conducts initial processing, the reservoir layer is typically defined by some nonlinear dynamic nodes with random and fixed connections and the readout layer recombines signals from the reservoir layer to achieve the desired target, as illustrated in Fig.~\ref{fig1}a. Compared to conventional photonic neural networks (PNN), photonic RC has some intrinsic advantages. In conventional PNN, precisely tuning the optical connections between different layers presents a significant challenge due to the manufacturing variations and the dynamics of thermal cross-talk~\cite{xu2022self}. Additionally, training PNNs is more difficult compared to training electronic ones because obtaining the gradient function of optical elements is complicated~\cite{pai2023experimentally}. In the photonic RC framework, the optical reservoir layer does not need to be trained and only the readout layer is required to be trained. The simple training procedure of photonic RC significantly reduces training time and power consumption.

 In traditional photonic RCs, optoelectronic components utilized for realizing nonlinear nodes play a crucial role. Since it is a challenge to realize optical nonlinearities efficiently, a widely adopted photonic RC structure leverages a single nonlinear node with time-delayed feedback to generate virtual nodes within the time domain~\cite{vinckier2015high,vatin2020experimental}. Despite its compact realization, this type of photonic RC is inherently slowed down due to time multiplexing. The time multiplexing method leads to a compromise between the number of neural nodes and processing speed. An alternative method to implement photonic RC uses spatially distributed nonlinear nodes~\cite{denis2018all,nakajima2021scalable, vandoorne2014experimental}. The scalability of such RC system is constrained by building up recurrent interconnections between the photonic nonlinear nodes, leading to a trade-off between the number of photonic neuron nodes and the overall system size. Furthermore, in most photonic RC systems, computational speed is constrained by the bandwidth of optoelectronic nonlinear components, such as those in ~\cite{vinckier2015high,lupo2023deep,shen2023deep}. Consequently, ultrafast and compact photonic RCs have not yet been achieved experimentally.

\definecolor{mycolor_0}{RGB}{214,213,193}
\definecolor{mycolor_1}{RGB}{250,250,246}
\definecolor{mycolor_2}{RGB}{236,235,226}

\begin{table}[h]
\caption{Performance comparison with previous photonic computing frameworks}
\begin{tabular}{cccccccc}
\hline

\rowcolor{mycolor_0}\multicolumn{2}{c}{\makecell{\textbf{Photonic Compu-}\\\textbf{ting Framework}}}&\makecell{\textbf{ Size }\\  \textbf{(mm$^2$)} } &\makecell{\textbf{Operation}\\\textbf{Speed\footnotemark[1]}\\\textbf{(GHz)} }&\makecell{\textbf{Inte-}\\\textbf{gration}}& \makecell{\textbf{Energy} \\ \textbf{Consump-}\\ \textbf{tion\footnotemark[2]}} & \makecell{\textbf{Metapa-}\\\textbf{rameters\footnotemark[3]}} & \makecell{\textbf{Computa-}\\\textbf{tional Density}\\\textbf{(TOPS/mm$^2$)}}\\
\hline
\rowcolor{mycolor_1}& ~\cite{vandoorne2014experimental} & 16  & 0.125$\sim$12.5&YES& without & without& N/A\footnotemark[4]\\

\rowcolor{mycolor_1}Photonic& ~\cite{vinckier2015high}& fibers &9 $\times$ 10$^{-4}$ &NO&with& with & N/A\\

\rowcolor{mycolor_1}  reservoir& ~\cite{nakajima2021scalable}  & 1320 & 0.06 &YES& with & with&0.016\\
\rowcolor{mycolor_1}computing &~\cite{lupo2023deep} &fibers& 0.01&NO&with &with& N/A\\
\rowcolor{mycolor_1}& ~\cite{shen2023deep} & fibers & 0.25 &NO&with & with& N/A\\

\hline
\rowcolor{mycolor_2}Photonic& ~\cite{feldmann2021parallel} & 3.63 & 12 &YES&with & N/A & 1.2\\
\rowcolor{mycolor_2}neural& ~\cite{ashtiani2022chip}  & 9.3& 21.7&YES& with & N/A &3.5\\
\rowcolor{mycolor_2}network& ~\cite{bai2023microcomb} & 0.131 & 17 &YES&with & N/A & 1.04\\
\hline

\rowcolor{mycolor_1}\multicolumn{2}{c}{\textbf{This work}} & \textbf{2}&\textbf{60}&\textbf{YES}&\textbf{without} & \textbf{without} & \textbf{103}\\
\hline
\end{tabular}
\label{table1}
\footnotetext[1]{Operation speed means the prediction or classification speed in photonic RC, and signal processing speed in photonic neural network.}
\footnotetext[2]{Energy for generating and detecting optical signals is not included here.}
\footnotetext[3]{Metaparameters are the parameters that need to be adjusted in the photonic reservoir layer.}
\footnotetext[4]{N/A represents no available data.}
\end{table}

Here, we propose and experimentally demonstrate a novel photonic RC system, leveraging a new computing framework called next-generation RC (NG-RC)~\cite{gauthier2021next}. NG-RC eliminates the need for an actual recurrent neural network to build connections; Instead, it directly calculates polynomial terms from time-delayed inputs. This framework allows us to realize nonlinearity in RC without requiring photonic nonlinear devices, thus eliminating speed limitations (except for limitations imposed by optical signal generation and detection) and to realize computing capability equivalent to recurrent neural networks without requiring physical recurrent connections. These features make our system to be high speed and highly scalable. Although recent studies have demonstrated the implementation of photonic NG-RC in free space and optical fibers~\cite{wang2024optical,cox2024photonic}, our system is the first time to achieve an integrated photonic NG-RC on a silicon chip with the fastest operation speed. Our structure is constructed with a compact star coupler and delay lines integrated on a silicon photonic chip with a compact area of 2 mm$^2$. Our novel photonic RC architecture exhibits significant advantages compared to previous optical computing systems. 

Table~\ref{table1} is a detailed performance comparison among previous photonic computing frameworks in terms of the footprint, operation speed, integration capability, power consumption, whether it requires metaparamters, and computational density. In addition to these advantages, our photonic RC also inherits the benefits of NG-RC and has the advantages of requiring no random interconnections, fewer delay lines, fewer metaparameters, and providing interpretable results, compared to conventional RCs. As a result, our system only requires a simple structure to realize high computing performance and has remarkable fabrication error tolerance, rendering it highly suitable for large-scale production. Most importantly, our photonic RC chip only consists of passive components, including a star coupler and on-chip delay lines, leading to no energy consumption and speed limitations (except for optical signal generation and detection). Our system experimentally realizes 103 TOPS mm$^{-2}$ computing density and computing rate of 60 Gbaud. To the best of our knowledge, this is the first time experimental demonstration of integrated photonic RC capable of prediction or classification with such a high speed. The speed can be further improved if using higher-speed optical modulators and photodetectors for optical signaling.

\section{Results}\label{sec2}

\subsection{Principle}\label{subsec2_1}
The core idea behind physical RC is designing and utilizing a dynamic system as a reservoir to map input data into higher dimensions~\cite{appeltant2011information}. Recent researches reveal that NG-RC can achieve comparable performance with optimized RC using no random interconnections, fewer metaparameters, shorter training data sets and training time~\cite{bollt2021explaining}. As shown in Fig.~\ref{fig1}b, the NG-RC functions equivalently to conventional RC but it does not require building a complex recurrent NN with nonlinear nodes. Instead, it is simply a concatenation of three vectors: a constant, a linear vector representing the input signal, and a vector representing the polynomial terms of the input signal, to realize a powerful universal approximation i.e.

\begin{equation}
O_{total} = C \oplus O_{linear} \oplus O_{nonlinear}\label{eq_1}
\end{equation}
where $\oplus$ represents the vector concatenation operation, $C$ is a constant, $O_{linear}$ is the linear vector, and $O_{nonlinear}$ is the nonlinear feature vector. The linear vector $O_{linear}$ is $X = [x_1, x_2,...,x_n]^T$. While it is flexible to select the nonlinear feature vector, the quadratic polynomial feature vector has been demonstrated good prediction capabilities~\cite{gauthier2021next}. Therefore nonlinear feature vector  $O_{nonlinear}$ is given by
\begin{equation}
O_{nonlinear} = O_{linear} \otimes O_{linear}\label{eq_2}
\end{equation}
where $\otimes$ represents the outer product operation. The final result $Y$ is derived by passing the $O_{total}$ through a simple trained readout layer. Notably, the NG-RC model has fewer parameters to be trained while achieving a high accuracy level comparable to those of traditional RC methods.

Fig.~\ref{fig1}c illustrates our proposed photonic RC structure based on NG-RC. The input data $X$ is encoded onto the amplitude of a laser, while $C$ represents the unmodulated laser. $X$ and $C$ are then sent to the reservoir, which consists of delay lines and a star coupler. The neighboring delay line introduces a time delay of $\Delta t$, which is equivalent to the delay caused by one symbol of the input signal. These delay lines ensure that input data from different times simultaneously reach the star coupler. Then, the evolution equation within the star coupler is expressed as
\begin{equation}
Y_{star} = W_{star}(C \oplus X) \label{eq_3}
\end{equation}
Here, $W_{star}$ denotes an $m \times (n+1)$ complex matrix, representing the transfer function of the star coupler, where $n$ is the number of delay lines and $m$ is the number of the output ports of the star coupler. The output dimension of our photonic RC depends on the input data dimension and their quadratic polynomial terms. And output $Y_{star}$ is a complex vector that characterizes the light after passing through the star coupler. The outputs of the star coupler are detected by photodiodes, which apply a quadratic transformation to the signal represented by $Y_{star}$. The equation of photodiodes can be described as
\begin{equation}
Y_{out} = |Y_{star}|^2 = |W_{star}(C \oplus X)|^2 \label{eq_4}
\end{equation}
Subsequently, the desired target is obtained by passing $Y_{out}$ through a straightforward trained readout layer.

The photodiode is the key component in generating the polynomial nonlinearity required by the NC-RC. The output of the $i_{th}$ photodiode is given by
\begin{equation}
\begin{aligned}
Y_{out,i} &= |W_{star,i}(C \oplus X)|^2  \\
&= c_i + \sum_{k=1}^{n}a_{i,k}x_{k} + \sum_{p=1}^{n}\left(\sum_{q=p}^{n}b_{i,p,q}x_px_q\right) \label{eq_5}
\end{aligned}
\end{equation}
where $c_i$, $a_{i,k}$, and $b_{i,p,q}$ are constants, which are determined by the structure of the star coupler. The Eq.~\eqref{eq_5} comprises constant $c_i$, the linear part $\sum_{k=1}^{n}a_{i,k}x_{k}$, and the quadratic polynomial $\sum_{p=1}^{n}\left(\sum_{q=p}^{n}b_{i,p,q}x_px_q\right)$. Therefore, the response of our reservoir RC is the same as the output of the NG-RC $O_{total}$. 

Importantly, the actual values of the $c_i$, $a_{i,k}$, and $b_{i,p,q}$ do not affect the final computing output because the final readout layer can adjust them to the optimal values through training. This feature makes our system have high fabrication error tolerance. Following the general theory of universal approximations, the dimension of input data is ideally considered to be infinitely large~\cite{gonon2019reservoir}. However, empirical observations reveal that the Volterra series converges rapidly, even when $n$ assumes small values~\cite{gauthier2021next}. Therefore, the number of delay lines does not need to be large and here we use eight delay lines for demonstration, leading to a small footprint of our photonic RC. Meanwhile, there is no speed limitation and power consumption in our passive photonic RC. We experimentally demonstrate a photonic RC at 60 Gbaud line rate within a compact area of 2 mm$^2$.

\subsection{System and Experimental setup}\label{subsec2_2}

\begin{figure*}[htbp]
\centerline{\includegraphics[scale=0.8]{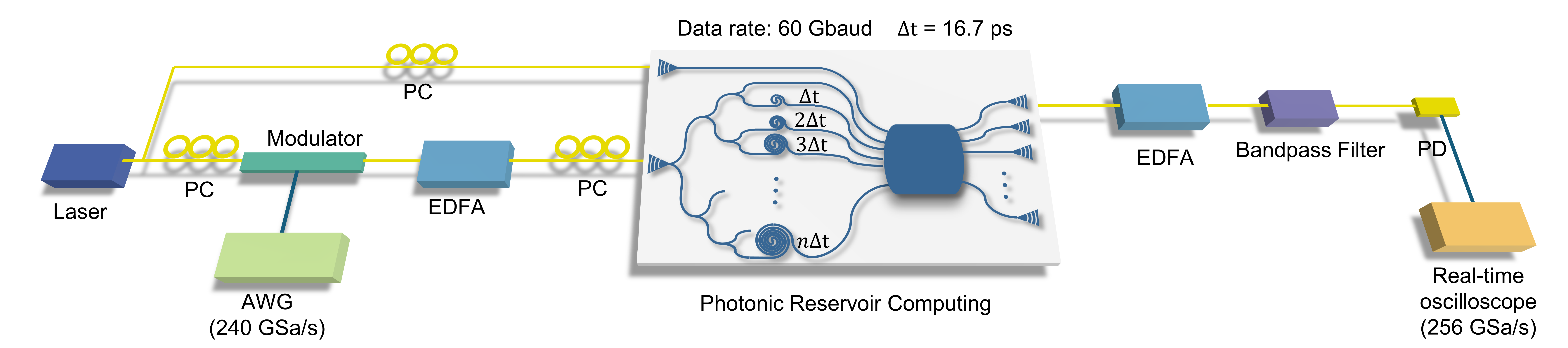}}
\caption{\textbf{Experimental setup}. PC, polarization controller; AWG, arbitrary waveform generator; EDFA, erbium-doped fiber amplifier; PD, photodiode. The unmodulated light and modulated signal as the constant and input data are injected into the photonic reservoir computing respectively.}
\label{fig2}
\end{figure*}

Fig.~\ref{fig2} shows the schematic of the experimental setup. We employ a continuous wave laser operating at 1550 nm, serving as our light source. Then, the light is split into two equal parts by a 50/50 coupler, one is coupled to the photonic chip as a constant reference $C$, and the other undergoes a thin-film lithium niobate intensity modulator (LIOBATE, LB4C6PSBM63, 40 GHz bandwidth, RF $V_{\pi} \approx$ 3 V). The input signal, generated by an arbitrary waveform generator (KEYSIGHT, M8199A, 256 GSa/s), is encoded onto the amplitude domain of light by the modulator. To achieve a 60 Gbaud line rate, we adjust the sample rate of the arbitrary waveform generator to 240 GSa/s. The modulator is configured in a push-pull arrangement and biased at the null point, corresponding to the minimum light intensity. The voltage applied to the modulator is chosen to  $|V| < V_{\pi}/4$, such that the input data can be linearly modulated onto the laser's amplitude. The modulated light is then amplified by an erbium-doped fiber amplifier (EDFA) before being injected into the photonic chip. The photonic chip, fabricated on a silicon-on-insulator (SOI) platform, consists of a star coupler and several integrated delay lines. The microscope photo and detailed design of this chip are shown in Supplementary Note 1. The signal is evenly divided into eight parts by on-chip couplers. Each part experiences a different delay through different delay lines. In our experiment, the neighboring delay line realizes $\Delta t = $ 16.7 ps delay. After passing through a compact star coupler, the signal is amplified again by another EDFA and filtered using a bandpass filter. Then, the signal is detected by an ultrafast photodetector (COHERENT, model XPDV3120R-VM-FA, with a 70 GHz bandwidth). The measured signal is digitized and recorded using an oscilloscope (KEYSIGHT, model UXR0592AP, with a 59 GHz bandwidth and 256 GSa/s sampling rate). Finally, the readout layer is trained at a digital processor.

\subsection{Experimental results}\label{sec2_3}

Here we present the experimental results for different benchmark tasks widely used in the reservoir computing framework. The prediction and classification speed of our photonic RC is 60 Gbaud in these experiments. 

\subsubsection{Lorenz system}\label{subsec2_3_1}

\begin{figure}[h]
\centerline{\includegraphics[scale=0.8]{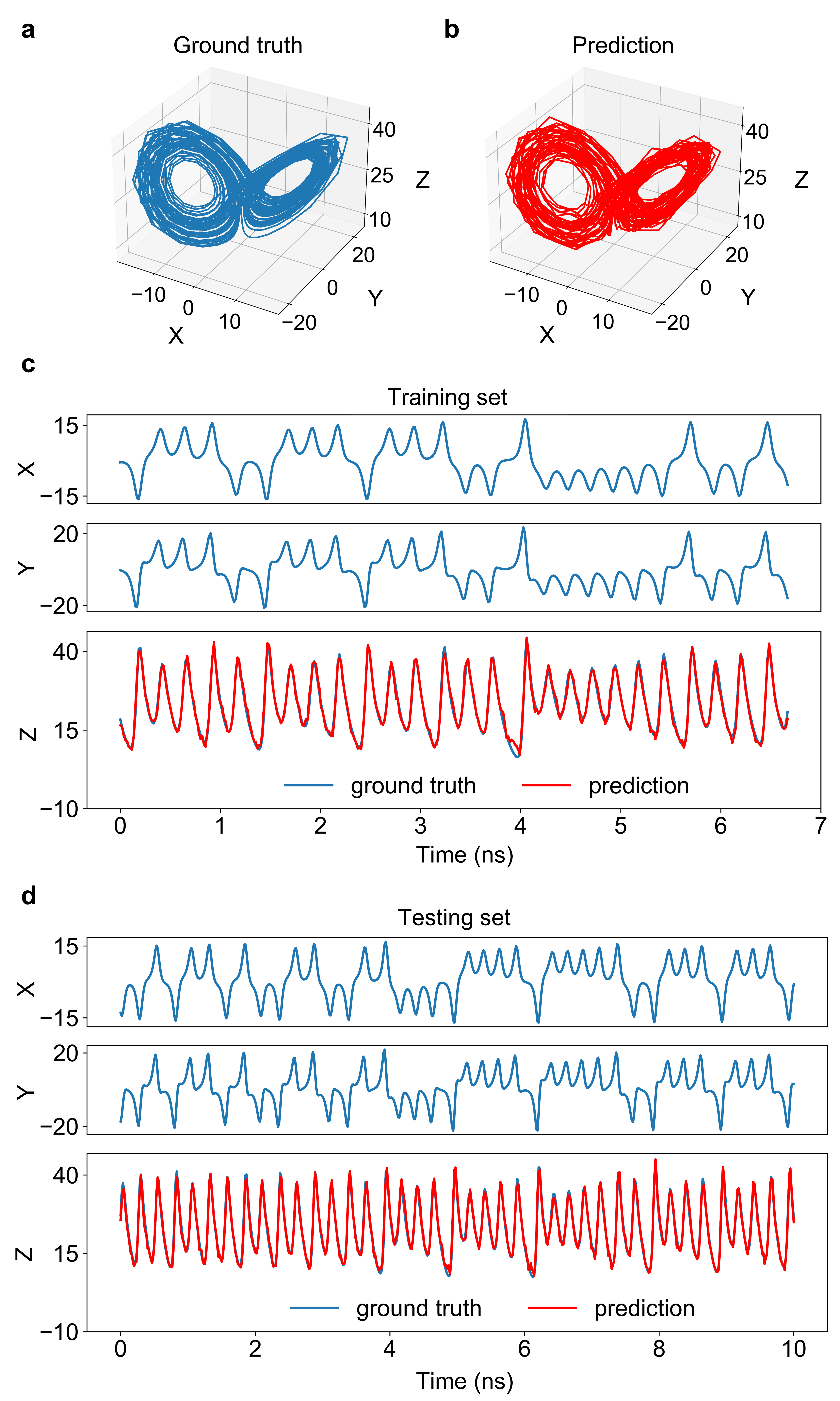}}
\caption{\textbf{Experiment result of Lorenz. a,} The ground truth three-dimensional diagram of the Lorenz system. \textbf{b,} The three-dimensional prediction result of our photonic system. \textbf{c,} The ground truth (blue) and prediction (red) variables of Lorenz during the training phase. \textbf{d,} The ground truth (blue) and prediction (red) variables of Lorenz during the testing phase. The NMSE of z variable between ground truth and prediction is $1.43 \times 10^{-2}$.}
\label{fig3}
\end{figure}

The first task is to use our system to predict chaotic sequences. Here we generate training and testing data using the Lorenz63 system, which is a simplified mathematical model developed by Lorenz in 1963 to study weather systems~\cite{lorenz1963deterministic}. The system consists of three ordinary differential equations representing the evolution of three variables: $x$, $y$, and $z$. The equations are given by
\begin{equation}
\begin{aligned}
\dot{x} &= 10(y-x),\\
\dot{y} &= x(28-z)-y,\\
\dot{z} &= xy - 8z/3,\label{eq_8}
\end{aligned}
\end{equation}
The Lorenz63 system serves as a paradigmatic example of deterministic chaos, as shown in Fig.~\ref{fig3}a. This nonlinear dynamic system can exhibit complex properties, including sensitivity to initial conditions (butterfly effect), and the existence of attractors with intricate geometric structures~\cite{gauthier2021next}.

Here, the aim of this task is to infer the next-step-ahead value of  $z$, from $x$ and $y$. This task holds significance for scenarios where acquiring precise data on a dynamical variable is feasible in controlled laboratory conditions. In such field settings, observable sensory data are utilized to estimate the absent information. Fig.~\ref{fig3}c shows the short training dataset for our photonic RC, comprising 400 points. The Tikhonov regularization is employed to train the readout layer. During the testing phase, $x$ and $y$ components are utilized to predict the next-step-ahead $z$ component using the following data set of 600 points. Fig.~\ref{fig3}b shows the 3D prediction result, which matches well with the ground truth. The testing results are depicted in Fig.~\ref{fig3}d. Normalized Mean Square Error (NMSE) is used to evaluate the prediction accuracy, which is defined as follows:
\begin{equation}
NMSE = \frac{\sum_n\left[ z^*(n) - z(n)\right]^2}{\sum_n\left\{z(n)-\frac{\sum_n[z(n)]}{n}\right\}^2}
\label{eq_10}
\end{equation}
The variable $z^*(n)$ represents the predicted outcome generated by our model. The NMSE of the testing $z$ is $1.43\times10^{-2}$, which implies the good prediction ability of our photonic RC.
\FloatBarrier

\subsubsection{NARMA10}\label{subsec2_3_2}

The second task is the nonlinear auto-regressive moving average (NARMA10) task, which serves as a commonly employed benchmark in the field of reservoir computing~\cite{jaeger2002adaptive}. The objective is to train a reservoir computer to emulate a nonlinear system of 10th order, with input $u(t)$ obtained randomly from a uniform distribution within the interval [0, 0.5]. The NARMA10 model is characterized by the recurrent equation
\begin{equation}
\begin{aligned}
y(n+1) &=  0.3\mu(n) + 0.05y(n)\left[\sum_{i=0}^9y(n-i)\right] \\
&+ 1.5\mu(n-9)\mu(n)+0.1\label{eq_9}
\end{aligned}
\end{equation}

\begin{figure}[htbp]
\centerline{\includegraphics[scale=0.8]{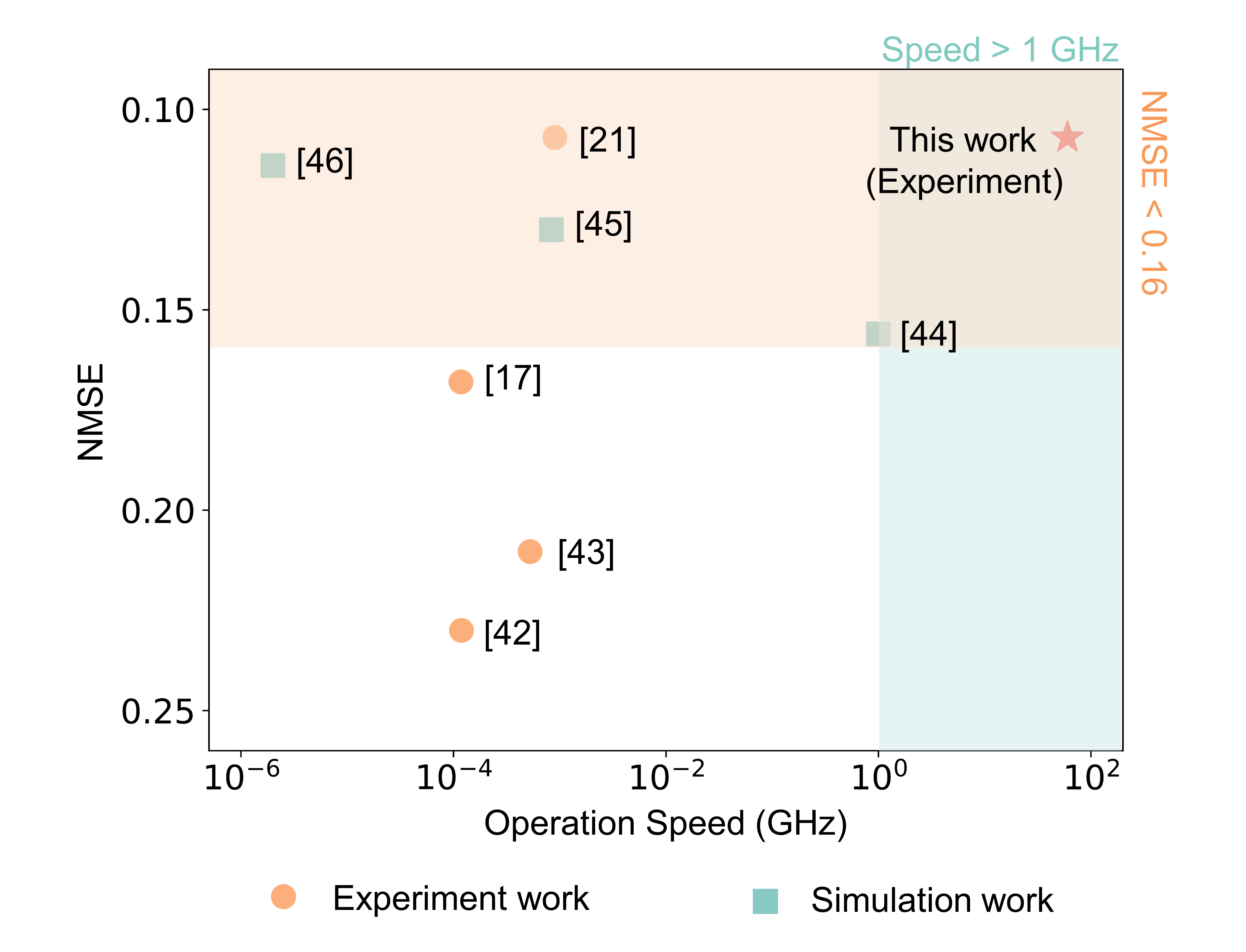}}
\caption{\textbf{The comparison of reported works.} The performance is evaluated with a focus on operation speed and the Normalized Mean Squared Error of the NARMA10 model. These works all have about 50 nonlinear nodes.}
\label{fig4}
\end{figure}

The prediction task relies on the historical input signal $\mu(n)$ and previous output values $y(n)$, indicating a significant need for memory. The objective of this benchmark task is to train a model to predict the next output value $y(n+1)$ based on input $\mu(n)$. We train our photonic RC module through a sequence spanning 1000 steps and test on the subsequent 1000 time steps. The NMSE in our experiment is 0.107 using $m = 45$ nodes. This NMSE is much lower than 0.16, which is the best performance achievable with no nonlinearity in the reservoir~\cite{appeltant2011information}. Up to now, the best experimental performance of this task reported in photonic RC with 50 nodes is NMSE = 0.107~\cite{vinckier2015high}, but with 0.9 MHz processing speed and large footprint. In Fig.~\ref{fig4}, we highlight the performance in this work in terms of operation speed and NMSE of NARMA10. The comparison results with existing experimental and simulation works show the superior performance of our structure~\cite{vinckier2015high,paquot2012optoelectronic,duport2016fully,huang2022multi,ren2024photonic,phang2023photonic,zheng2021parameters}.

\subsubsection{COVID-19}\label{subsec2_3_3}

\begin{figure}[htbp]
\centerline{\includegraphics[scale=0.8]{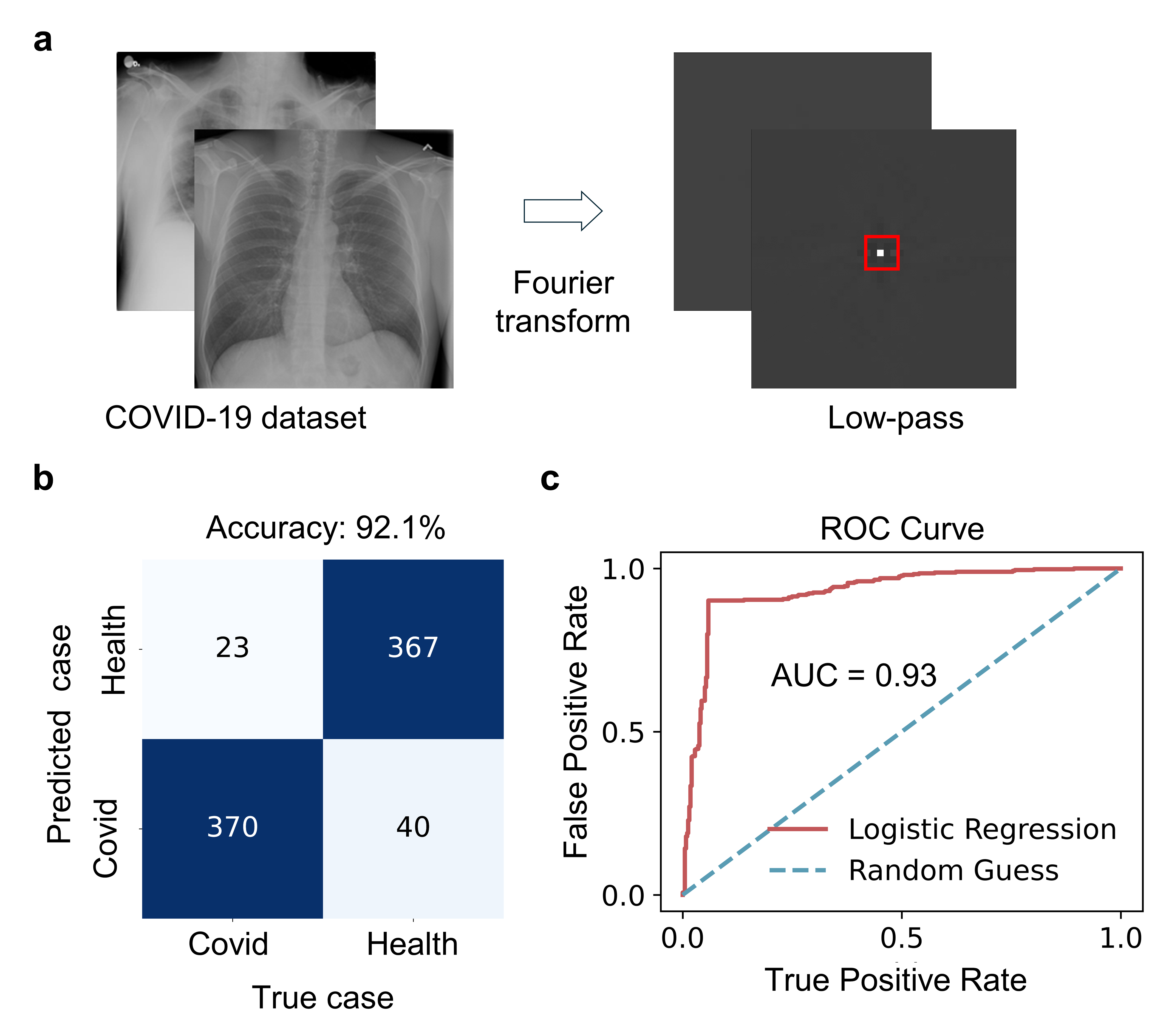}}
\caption{\textbf{Experiment results of Covid-19 classification. a,} The preprocessing of the Covid-19 images. \textbf{b,} The confusion matrices for Covid-19 classification. The accuracy of the Covid-19 classification is 92.1\% \textbf{c,} The receiver operating characteristic curve of Covid-19 classification. The area under the curve of this task is 0.93.}
\label{fig5}
\end{figure}

The implemented photonic RC can also be used for image classification. Here we demonstrate a two-class COVID-19 classification task~\cite{teugin2021scalable}. The dataset comprises 2,000 healthy X-ray samples and 2,000 COVID-19 samples and each sample has 299 $\times$ 299 pixels. We initially performed image preprocessing, including Fourier transform and low-pass filtering, to decrease input dimensions, as depicted in Fig.~\ref{fig5}a. The Fourier transform can be regarded as a form of unitary matrix operation~\cite{vadlamani2023transferable}. Therefore, these preprocessing steps can be accomplished using an integrated MZI mesh~\cite{perez2019scalable} or simply an optical lens. The real parts of the eight slowest spacial frequencies from these raw images are utilized to train and test our photonic RC. In the readout layer, logistic regression is used to perform classification. We employ 3,200 data samples as the training sets, and 800 data samples as the testing sets. The results are shown in Fig.~\ref{fig5}b-c. The accuracy of this classification task is 92.1\%, which is higher than the ones in the references~\cite{teugin2021scalable, momeni2022electromagnetic}. We also use receiver operating characteristic curve (ROC) to evaluate the performance of our system. The area under the curve (AUC) is 0.93, indicating the excellent classification performance of the photonic RC.

\section{Conclusion}\label{sec4}

In summary, we have experimentally implemented a computing engine on photonic integrated circuits operating at 60 Gbaud line rate and a computing density of 103 TOPS mm$^{-2}$ to realize next-generation RC. Our experiment results showed high performance in both prediction and classification tasks. Meanwhile, our photonic RC has higher fabrication error tolerance, fewer metaparameters, and greater interpretability than conventional photonic RC, making it easy to achieve large-scale industrial production. The computing engine consists of a straightforward small-footprint star coupler with on-chip delay lines, leading to compact size, no energy consumption, and no speed limitation. 

Moreover, the computational density of our chip is 103 TOPS mm$^{-2}$. The comprehensive analysis of computational density is shown in Supplementary Note 3. The operation speed of this computing engine can easily reach up to several hundreds of GHz by shortening the length of delay lines and using higher-speed optical modulators and photodetectors~\cite{wang2018integrated,lischke2021ultra}. The size of our system can be reduced to less than 1 mm$^2$ by optimizing the system layout. In this way, the computational density of this engine can be expanded to peta operations per second mm$^{-2}$.

Furthermore, the readout layer of our engine can be realized by on-chip modulators, which makes it possible to achieve a fully integrated photonic RC. Nowadays, the integration of high-speed modulators and PDs into a silicon photonic chip is achievable through commercial foundries. Our whole system, including encoding and detecting optical signals, can be implemented within a silicon photonic chip in the future. Thus, our computing engine opens a new route for the development of next-generation ultrafast photonic computing.


\bibliography{sn-bibliography}

\begin{thebibliography}{10}
\providecommand{\url}[1]{#1}
\csname url@samestyle\endcsname
\providecommand{\newblock}{\relax}
\providecommand{\bibinfo}[2]{#2}
\providecommand{\BIBentrySTDinterwordspacing}{\spaceskip=0pt\relax}
\providecommand{\BIBentryALTinterwordstretchfactor}{4}
\providecommand{\BIBentryALTinterwordspacing}{\spaceskip=\fontdimen2\font plus
\BIBentryALTinterwordstretchfactor\fontdimen3\font minus \fontdimen4\font\relax}
\providecommand{\BIBforeignlanguage}[2]{{%
\expandafter\ifx\csname l@#1\endcsname\relax
\typeout{** WARNING: IEEEtran.bst: No hyphenation pattern has been}%
\typeout{** loaded for the language `#1'. Using the pattern for}%
\typeout{** the default language instead.}%
\else
\language=\csname l@#1\endcsname
\fi
#2}}
\providecommand{\BIBdecl}{\relax}
\BIBdecl

\bibitem{furber2016large}
S.~Furber, ``Large-scale neuromorphic computing systems,'' \emph{Journal of neural engineering}, vol.~13, no.~5, p. 051001, 2016.

\bibitem{liu2024sora}
Y.~Liu, K.~Zhang, Y.~Li, Z.~Yan, C.~Gao, R.~Chen, Z.~Yuan, Y.~Huang, H.~Sun, J.~Gao \emph{et~al.}, ``Sora: A review on background, technology, limitations, and opportunities of large vision models,'' \emph{arXiv preprint arXiv:2402.17177}, 2024.

\bibitem{shen2017deep}
Y.~Shen, N.~C. Harris, S.~Skirlo, M.~Prabhu, T.~Baehr-Jones, M.~Hochberg, X.~Sun, S.~Zhao, H.~Larochelle, D.~Englund \emph{et~al.}, ``Deep learning with coherent nanophotonic circuits,'' \emph{Nature photonics}, vol.~11, no.~7, pp. 441--446, 2017.

\bibitem{lin2018all}
X.~Lin, Y.~Rivenson, N.~T. Yardimci, M.~Veli, Y.~Luo, M.~Jarrahi, and A.~Ozcan, ``All-optical machine learning using diffractive deep neural networks,'' \emph{Science}, vol. 361, no. 6406, pp. 1004--1008, 2018.

\bibitem{feldmann2019all}
J.~Feldmann, N.~Youngblood, C.~D. Wright, H.~Bhaskaran, and W.~H. Pernice, ``All-optical spiking neurosynaptic networks with self-learning capabilities,'' \emph{Nature}, vol. 569, no. 7755, pp. 208--214, 2019.

\bibitem{xu202111}
X.~Xu, M.~Tan, B.~Corcoran, J.~Wu, A.~Boes, T.~G. Nguyen, S.~T. Chu, B.~E. Little, D.~G. Hicks, R.~Morandotti \emph{et~al.}, ``11 tops photonic convolutional accelerator for optical neural networks,'' \emph{Nature}, vol. 589, no. 7840, pp. 44--51, 2021.

\bibitem{shastri2021photonics}
B.~J. Shastri, A.~N. Tait, T.~Ferreira~de Lima, W.~H. Pernice, H.~Bhaskaran, C.~D. Wright, and P.~R. Prucnal, ``Photonics for artificial intelligence and neuromorphic computing,'' \emph{Nature Photonics}, vol.~15, no.~2, pp. 102--114, 2021.

\bibitem{huang2021silicon}
C.~Huang, S.~Fujisawa, T.~F. de~Lima, A.~N. Tait, E.~C. Blow, Y.~Tian, S.~Bilodeau, A.~Jha, F.~Yaman, H.-T. Peng \emph{et~al.}, ``A silicon photonic--electronic neural network for fibre nonlinearity compensation,'' \emph{Nature Electronics}, vol.~4, no.~11, pp. 837--844, 2021.

\bibitem{feldmann2021parallel}
J.~Feldmann, N.~Youngblood, M.~Karpov, H.~Gehring, X.~Li, M.~Stappers, M.~Le~Gallo, X.~Fu, A.~Lukashchuk, A.~S. Raja \emph{et~al.}, ``Parallel convolutional processing using an integrated photonic tensor core,'' \emph{Nature}, vol. 589, no. 7840, pp. 52--58, 2021.

\bibitem{xu2022self}
X.~Xu, G.~Ren, T.~Feleppa, X.~Liu, A.~Boes, A.~Mitchell, and A.~J. Lowery, ``Self-calibrating programmable photonic integrated circuits,'' \emph{Nature Photonics}, vol.~16, no.~8, pp. 595--602, 2022.

\bibitem{wang2023image}
T.~Wang, M.~M. Sohoni, L.~G. Wright, M.~M. Stein, S.-Y. Ma, T.~Onodera, M.~G. Anderson, and P.~L. McMahon, ``Image sensing with multilayer nonlinear optical neural networks,'' \emph{Nature Photonics}, vol.~17, no.~5, pp. 408--415, 2023.

\bibitem{xu2024large}
Z.~Xu, T.~Zhou, M.~Ma, C.~Deng, Q.~Dai, and L.~Fang, ``Large-scale photonic chiplet taichi empowers 160-tops/w artificial general intelligence,'' \emph{Science}, vol. 384, no. 6692, pp. 202--209, 2024.

\bibitem{dong2023higher}
B.~Dong, S.~Aggarwal, W.~Zhou, U.~E. Ali, N.~Farmakidis, J.~S. Lee, Y.~He, X.~Li, D.-L. Kwong, C.~Wright \emph{et~al.}, ``Higher-dimensional processing using a photonic tensor core with continuous-time data,'' \emph{Nature Photonics}, vol.~17, no.~12, pp. 1080--1088, 2023.

\bibitem{ashtiani2022chip}
F.~Ashtiani, A.~J. Geers, and F.~Aflatouni, ``An on-chip photonic deep neural network for image classification,'' \emph{Nature}, vol. 606, no. 7914, pp. 501--506, 2022.

\bibitem{chen2023all}
Y.~Chen, M.~Nazhamaiti, H.~Xu, Y.~Meng, T.~Zhou, G.~Li, J.~Fan, Q.~Wei, J.~Wu, F.~Qiao \emph{et~al.}, ``All-analog photoelectronic chip for high-speed vision tasks,'' \emph{Nature}, vol. 623, no. 7985, pp. 48--57, 2023.

\bibitem{liu2022programmable}
C.~Liu, Q.~Ma, Z.~J. Luo, Q.~R. Hong, Q.~Xiao, H.~C. Zhang, L.~Miao, W.~M. Yu, Q.~Cheng, L.~Li \emph{et~al.}, ``A programmable diffractive deep neural network based on a digital-coding metasurface array,'' \emph{Nature Electronics}, vol.~5, no.~2, pp. 113--122, 2022.

\bibitem{paquot2012optoelectronic}
Y.~Paquot, F.~Duport, A.~Smerieri, J.~Dambre, B.~Schrauwen, M.~Haelterman, and S.~Massar, ``Optoelectronic reservoir computing,'' \emph{Scientific reports}, vol.~2, no.~1, p. 287, 2012.

\bibitem{larger2012photonic}
L.~Larger, M.~C. Soriano, D.~Brunner, L.~Appeltant, J.~M. Guti{\'e}rrez, L.~Pesquera, C.~R. Mirasso, and I.~Fischer, ``Photonic information processing beyond turing: an optoelectronic implementation of reservoir computing,'' \emph{Optics express}, vol.~20, no.~3, pp. 3241--3249, 2012.

\bibitem{brunner2013parallel}
D.~Brunner, M.~C. Soriano, C.~R. Mirasso, and I.~Fischer, ``Parallel photonic information processing at gigabyte per second data rates using transient states,'' \emph{Nature communications}, vol.~4, no.~1, p. 1364, 2013.

\bibitem{vandoorne2014experimental}
K.~Vandoorne, P.~Mechet, T.~Van~Vaerenbergh, M.~Fiers, G.~Morthier, D.~Verstraeten, B.~Schrauwen, J.~Dambre, and P.~Bienstman, ``Experimental demonstration of reservoir computing on a silicon photonics chip,'' \emph{Nature communications}, vol.~5, no.~1, p. 3541, 2014.

\bibitem{vinckier2015high}
Q.~Vinckier, F.~Duport, A.~Smerieri, K.~Vandoorne, P.~Bienstman, M.~Haelterman, and S.~Massar, ``High-performance photonic reservoir computer based on a coherently driven passive cavity,'' \emph{Optica}, vol.~2, no.~5, pp. 438--446, 2015.

\bibitem{larger2017high}
L.~Larger, A.~Bayl{\'o}n-Fuentes, R.~Martinenghi, V.~S. Udaltsov, Y.~K. Chembo, and M.~Jacquot, ``High-speed photonic reservoir computing using a time-delay-based architecture: Million words per second classification,'' \emph{Physical Review X}, vol.~7, no.~1, p. 011015, 2017.

\bibitem{antonik2019human}
P.~Antonik, N.~Marsal, D.~Brunner, and D.~Rontani, ``Human action recognition with a large-scale brain-inspired photonic computer,'' \emph{Nature Machine Intelligence}, vol.~1, no.~11, pp. 530--537, 2019.

\bibitem{rafayelyan2020large}
M.~Rafayelyan, J.~Dong, Y.~Tan, F.~Krzakala, and S.~Gigan, ``Large-scale optical reservoir computing for spatiotemporal chaotic systems prediction,'' \emph{Physical Review X}, vol.~10, no.~4, p. 041037, 2020.

\bibitem{sunada2021photonic}
S.~Sunada and A.~Uchida, ``Photonic neural field on a silicon chip: large-scale, high-speed neuro-inspired computing and sensing,'' \emph{Optica}, vol.~8, no.~11, pp. 1388--1396, 2021.

\bibitem{nakajima2021scalable}
M.~Nakajima, K.~Tanaka, and T.~Hashimoto, ``Scalable reservoir computing on coherent linear photonic processor,'' \emph{Communications Physics}, vol.~4, no.~1, p.~20, 2021.

\bibitem{lupo2023deep}
A.~Lupo, E.~Picco, M.~Zajnulina, and S.~Massar, ``Deep photonic reservoir computer based on frequency multiplexing with fully analog connection between layers,'' \emph{Optica}, vol.~10, no.~11, pp. 1478--1485, 2023.

\bibitem{shen2023deep}
Y.-W. Shen, R.-Q. Li, G.-T. Liu, J.~Yu, X.~He, L.~Yi, and C.~Wang, ``Deep photonic reservoir computing recurrent network,'' \emph{Optica}, vol.~10, no.~12, pp. 1745--1751, 2023.

\bibitem{yan2024emerging}
M.~Yan, C.~Huang, P.~Bienstman, P.~Tino, W.~Lin, and J.~Sun, ``Emerging opportunities and challenges for the future of reservoir computing,'' \emph{Nature Communications}, vol.~15, no.~1, p. 2056, 2024.

\bibitem{pai2023experimentally}
S.~Pai, Z.~Sun, T.~W. Hughes, T.~Park, B.~Bartlett, I.~A. Williamson, M.~Minkov, M.~Milanizadeh, N.~Abebe, F.~Morichetti \emph{et~al.}, ``Experimentally realized in situ backpropagation for deep learning in photonic neural networks,'' \emph{Science}, vol. 380, no. 6643, pp. 398--404, 2023.

\bibitem{vatin2020experimental}
J.~Vatin, D.~Rontani, and M.~Sciamanna, ``Experimental realization of dual task processing with a photonic reservoir computer,'' \emph{APL Photonics}, vol.~5, no.~8, 2020.

\bibitem{denis2018all}
F.~Denis-Le~Coarer, M.~Sciamanna, A.~Katumba, M.~Freiberger, J.~Dambre, P.~Bienstman, and D.~Rontani, ``All-optical reservoir computing on a photonic chip using silicon-based ring resonators,'' \emph{IEEE Journal of Selected Topics in Quantum Electronics}, vol.~24, no.~6, pp. 1--8, 2018.

\bibitem{bai2023microcomb}
B.~Bai, Q.~Yang, H.~Shu, L.~Chang, F.~Yang, B.~Shen, Z.~Tao, J.~Wang, S.~Xu, W.~Xie \emph{et~al.}, ``Microcomb-based integrated photonic processing unit,'' \emph{Nature Communications}, vol.~14, no.~1, p.~66, 2023.

\bibitem{gauthier2021next}
D.~J. Gauthier, E.~Bollt, A.~Griffith, and W.~A. Barbosa, ``Next generation reservoir computing,'' \emph{Nature communications}, vol.~12, no.~1, pp. 1--8, 2021.

\bibitem{wang2024optical}
H.~Wang, J.~Hu, Y.~Baek, K.~Tsuchiyama, M.~Joly, Q.~Liu, and S.~Gigan, ``Optical next generation reservoir computing,'' \emph{arXiv preprint arXiv:2404.07857}, 2024.

\bibitem{cox2024photonic}
N.~Cox, J.~Murray, J.~Hart, and B.~Redding, ``Photonic next-generation reservoir computer based on distributed feedback in optical fiber,'' \emph{arXiv preprint arXiv:2404.07116}, 2024.

\bibitem{appeltant2011information}
L.~Appeltant, M.~C. Soriano, G.~Van~der Sande, J.~Danckaert, S.~Massar, J.~Dambre, B.~Schrauwen, C.~R. Mirasso, and I.~Fischer, ``Information processing using a single dynamical node as complex system,'' \emph{Nature communications}, vol.~2, no.~1, p. 468, 2011.

\bibitem{bollt2021explaining}
E.~Bollt, ``On explaining the surprising success of reservoir computing forecaster of chaos? the universal machine learning dynamical system with contrast to var and dmd,'' \emph{Chaos: An Interdisciplinary Journal of Nonlinear Science}, vol.~31, no.~1, 2021.

\bibitem{gonon2019reservoir}
L.~Gonon and J.-P. Ortega, ``Reservoir computing universality with stochastic inputs,'' \emph{IEEE transactions on neural networks and learning systems}, vol.~31, no.~1, pp. 100--112, 2019.

\bibitem{lorenz1963deterministic}
E.~N. Lorenz, ``Deterministic nonperiodic flow,'' \emph{Journal of atmospheric sciences}, vol.~20, no.~2, pp. 130--141, 1963.

\bibitem{jaeger2002adaptive}
H.~Jaeger, ``Adaptive nonlinear system identification with echo state networks,'' \emph{Advances in neural information processing systems}, vol.~15, 2002.

\bibitem{duport2016fully}
F.~Duport, A.~Smerieri, A.~Akrout, M.~Haelterman, and S.~Massar, ``Fully analogue photonic reservoir computer,'' \emph{Scientific reports}, vol.~6, no.~1, p. 22381, 2016.

\bibitem{huang2022multi}
L.~Huang and J.~Yao, ``Multi-task photonic time-delay reservoir computing based on polarization modulation,'' \emph{Optics Letters}, vol.~47, no.~24, pp. 6464--6467, 2022.

\bibitem{ren2024photonic}
H.~Ren, Y.~Li, M.~Li, M.~Gao, J.~Lu, C.-L. Zou, C.-H. Dong, P.~Yu, X.~Yang, and Q.~Xuan, ``Photonic time-delayed reservoir computing based on series-coupled microring resonators with high memory capacity,'' \emph{Optics Express}, vol.~32, no.~7, pp. 11\,202--11\,220, 2024.

\bibitem{phang2023photonic}
S.~Phang, ``Photonic reservoir computing enabled by stimulated brillouin scattering,'' \emph{Optics Express}, vol.~31, no.~13, pp. 22\,061--22\,074, 2023.

\bibitem{zheng2021parameters}
T.~Zheng, W.~Yang, J.~Sun, X.~Xiong, Z.~Li, and X.~Zou, ``Parameters optimization method for the time-delayed reservoir computing with a nonlinear duffing mechanical oscillator,'' \emph{Scientific Reports}, vol.~11, no.~1, p. 997, 2021.

\bibitem{teugin2021scalable}
U.~Te{\u{g}}in, M.~Y{\i}ld{\i}r{\i}m, {\.I}.~O{\u{g}}uz, C.~Moser, and D.~Psaltis, ``Scalable optical learning operator,'' \emph{Nature Computational Science}, vol.~1, no.~8, pp. 542--549, 2021.

\bibitem{vadlamani2023transferable}
S.~K. Vadlamani, D.~Englund, and R.~Hamerly, ``Transferable learning on analog hardware,'' \emph{Science Advances}, vol.~9, no.~28, p. eadh3436, 2023.

\bibitem{perez2019scalable}
D.~P{\'e}rez and J.~Capmany, ``Scalable analysis for arbitrary photonic integrated waveguide meshes,'' \emph{Optica}, vol.~6, no.~1, pp. 19--27, 2019.

\bibitem{momeni2022electromagnetic}
A.~Momeni and R.~Fleury, ``Electromagnetic wave-based extreme deep learning with nonlinear time-floquet entanglement,'' \emph{Nature Communications}, vol.~13, no.~1, p. 2651, 2022.

\bibitem{wang2018integrated}
C.~Wang, M.~Zhang, X.~Chen, M.~Bertrand, A.~Shams-Ansari, S.~Chandrasekhar, P.~Winzer, and M.~Lon{\v{c}}ar, ``Integrated lithium niobate electro-optic modulators operating at cmos-compatible voltages,'' \emph{Nature}, vol. 562, no. 7725, pp. 101--104, 2018.

\bibitem{lischke2021ultra}
S.~Lischke, A.~Peczek, J.~Morgan, K.~Sun, D.~Steckler, Y.~Yamamoto, F.~Kornd{\"o}rfer, C.~Mai, S.~Marschmeyer, M.~Fraschke \emph{et~al.}, ``Ultra-fast germanium photodiode with 3-db bandwidth of 265 ghz,'' \emph{Nature Photonics}, vol.~15, no.~12, pp. 925--931, 2021.

\bibitem{zhou2021large}
T.~Zhou, X.~Lin, J.~Wu, Y.~Chen, H.~Xie, Y.~Li, J.~Fan, H.~Wu, L.~Fang, and Q.~Dai, ``Large-scale neuromorphic optoelectronic computing with a reconfigurable diffractive processing unit,'' \emph{Nature Photonics}, vol.~15, no.~5, pp. 367--373, 2021.

\end{thebibliography}

\section*{Acknowledgements}\label{sec5}

This work was supported by ITF ITS/237/22, ITS/226/21FP, RGC YCRF C1002-22Y, RNE-p4-22 of the Shun Hing Institute of Advanced Engineering, NSFC/RGC Joint Research Scheme N CUHK444/22, and CUHK Direct Grant 170257018, 4055143.

\section*{Author contributions}\label{sec6}

D.W. and C.H. conceived the ideas. D.W. and G.H. designed and simulated the system structure. D.W. designed the experiment. D.W. and Y.N. completed the experimental data collection and analyzed the results. D.W., Y.N. and C.H. wrote the manuscript. H.K.T. provided experimental advice and revised the manuscript. C.H. supervised the research and contributed to the general concept and interpretation of the results. All the authors contributed to the manuscript.
\newpage
\section*{\Large{A 103-TOPS/mm$^2$ Integrated Photonic Computing Engine Enabling Next-Generation Reservoir Computing: supplemental document}}

\vspace{0.5cm}
\large
Supplementary Note 1. \textbf{Device structure}\\
Supplementary Note 2. \textbf{Data processing for Lorenz and NARMA10 task} \\
Supplementary Note 3. \textbf{Computational density}
\vspace{0.5cm}

\newcommand{\mysection}[2]{
  \setcounter{section}{#1}
  \addtocounter{section}{-1}
  \section{ #2}
}

\mysection{1}{Device structure}

\begin{figure}[htbp]
\centerline{\includegraphics{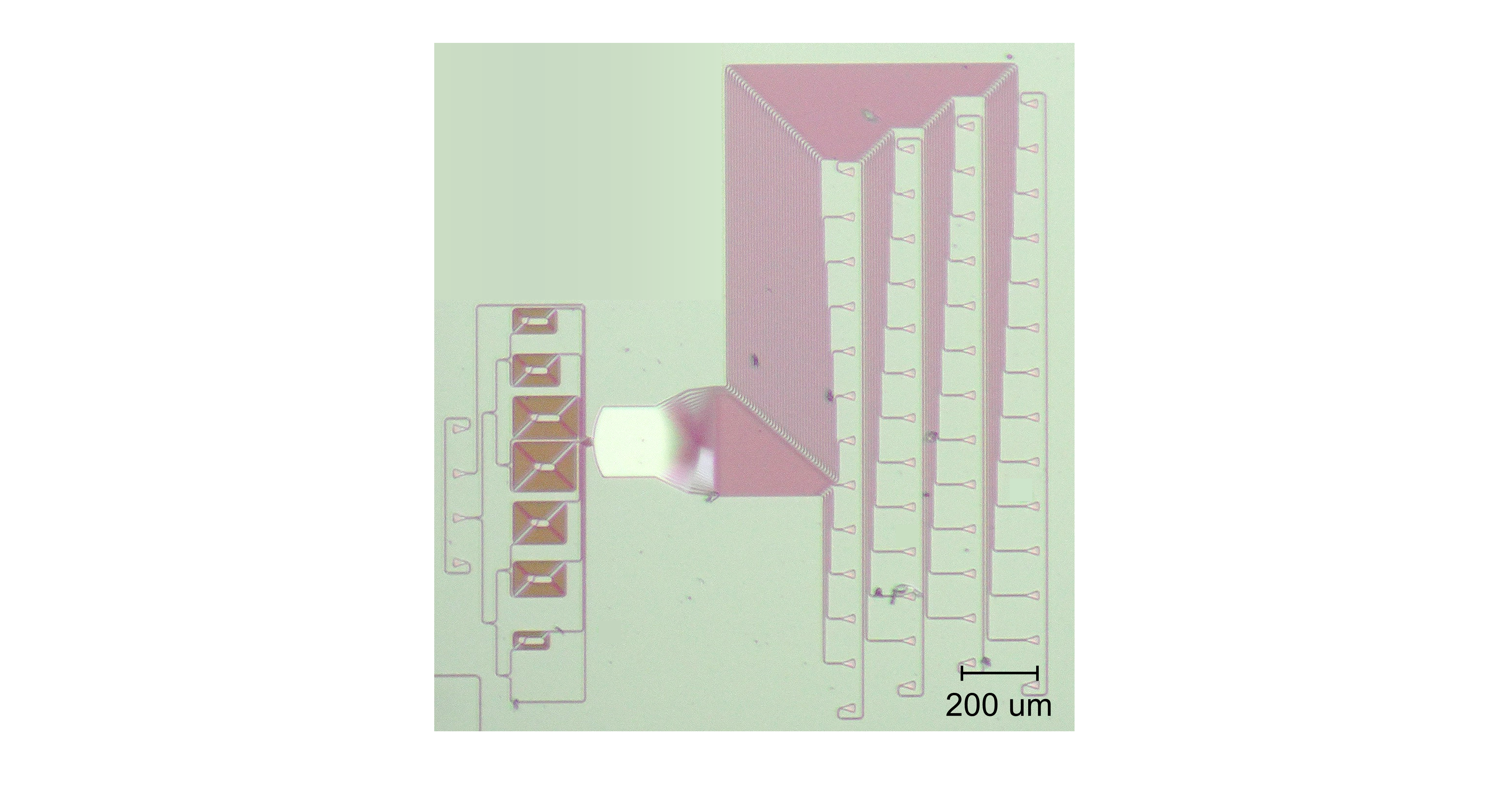}}
\caption{\textbf{Device structure.} The microscope photo of our photonic next-generation reservoir computing structure.}
\label{supplementary_fig1}
\end{figure}

Our photonic reservoir computing structure is fabricated on a silicon-on-insulator (SOI) platform by Applied Nanotools Inc. (ANT), as depicted in Figure \ref{supplementary_fig1}. The structure is composed of a star coupler and several integrated delay lines. The star coupler, with an area of 200 $\mu$m $\times$ 200 $\mu$m, has nine input ports and forty-five output ports. The quantity of output ports is determined by the input data and its corresponding quadratic polynomial feature vector. Our experiment has nine input data (including constant) and thirty-six corresponding quadratic polynomials. Thus, the star coupler has forty-five output ports in total. 

\begin{figure}[htbp]
\centerline{\includegraphics{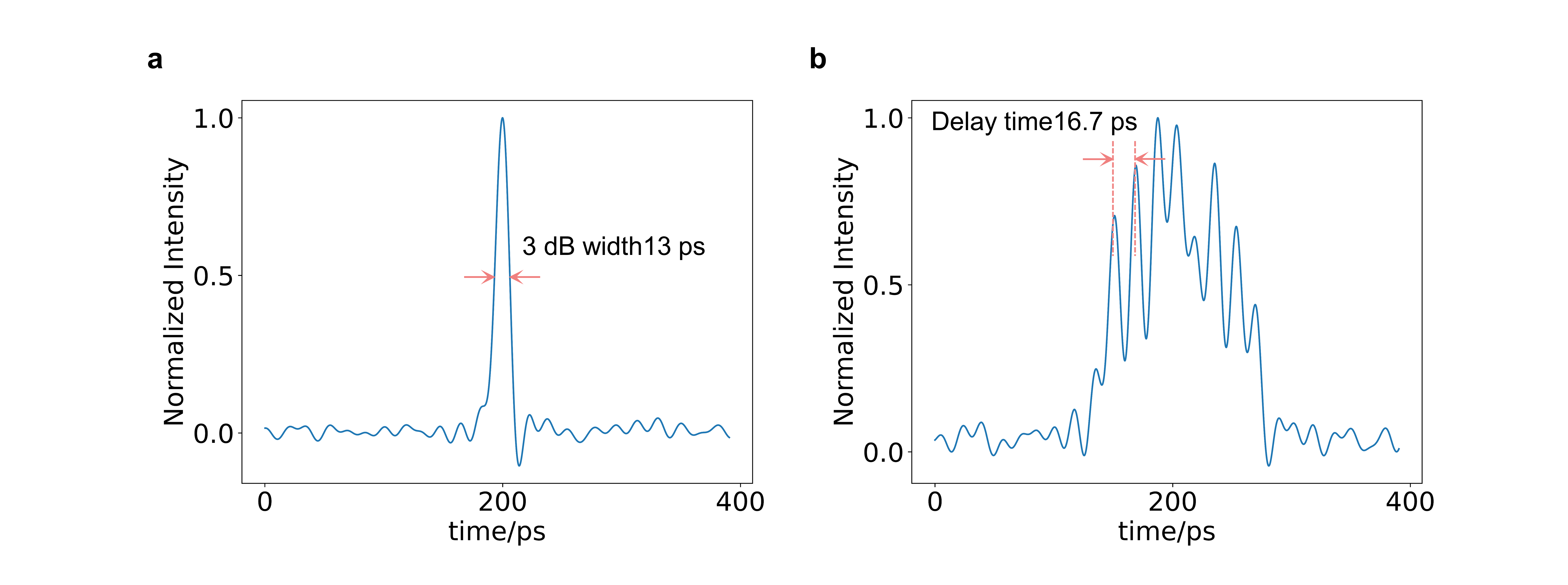}}
\caption{\textbf{Delay line test.} \textbf{a} The optical pulse is used to test the delay time of our delay line. The 3 dB width of this pulse is 13 ps. \textbf{b} The output optical pulse after the tested optical pulse passes through our structure. The neighboring delay line realizes a 16.7 ps delay.}
\label{supplementary_fig2}
\end{figure}

The on-chip delay lines in our system are composed of three layers: a 2.2-$\mu$m-thick oxide (SiO$_2$) cladding layer, a 220-nm-thick silicon (Si) layer, and a 2-$\mu$m-thick buried thermal oxide ((SiO$_2$) layer. The width of the silicon layer is 500 nm in our design. The group index of these delay lines is approximately 4.24 at $\lambda$ = 1550 nm by numerical simulation. To achieve a delay of 16.7 ps in our design, the length difference of the neighboring delay lines is set as 1.18 mm. We employ a 13-ps width optical pulse to evaluate the delay time of different delay lines, as depicted in Figure \ref{supplementary_fig2}a. Figure \ref{supplementary_fig2}b displays the output optical pulse after the tested optical pulse passes through our structure. The adjacent delay line realizes $\Delta$ t = 16.7 ps.

\mysection{2}{Data processing for Lorenz and NARMA10 task}

In the Lorenz task, variables x, and y are used to infer the next-step-ahead prediction of the variable z. We concatenate the time series of x and y at time points t, t-5, t-10, and t-15, and use them as inputs to our system. This task aims to predict z at time t+1, same as the approach in Ref.~\cite{gauthier2021next}.

In the NARMA10 task, we use the variable $\mu$(n) to predict y(n+1). We concatenate $\mu$ at time points n, n-1, n-2, n-3, n-9, n-10, n-11, and n-12, and use these as inputs to our system. This task aims to predict y at time n+1.

\mysection{3}{Computational density}
Our system comprises nine input ports (N = 9) and forty-five output ports (M = 45). The process can be simplified as an N $\times$ M complex matrix processing and an M-complex quadratic operation. As noted in ~\cite{xu2024large, zhou2021large}, each complete multiplication involves 4 real multiplications and  2 real summations, while each complex summation involves 2 real summations. An N $\times$ M complex matrix processing encompasses NM complex multiplications and (N-1)M complex summations. Therefore, the total operation of a N $\times$ M complex matrix processing has 6NM + 2(N-1)M computational operations. An M-complex quadratic operation has M-complex multiplications, leading to 6M computational operations. Our system operates at a line rate of 60 Gbaud, resulting in a total optical computational operation of 205 TOPS. The whole system occupies a space of  2 mm$^2$, yielding a computational density of 103 TOPS/mm$^2$. As shown in  Figure \ref{fig1}, the system layout is currently not optimized. By refining the system layout, the footprint could potentially be reduced to less than 1 mm$^2$. Therefore, computational density can easily reach up to 205 TOPS/mm$^2$ only by optimizing the system layout.

\end{document}